\documentclass{PoS}

\newcommand{\be}{\begin{equation}}
\newcommand{\ee}{\end{equation}}
\newcommand{\bea}{\begin{eqnarray}}
\newcommand{\eea}{\end{eqnarray}}

\title{Thermodynamics of $SU(N)$ gauge theories in $2+1$ dimensions in the      
$T <\ T_c$ regime}

\ShortTitle{Thermodynamics of $SU(N)$ gauge theories in $2+1$ dimensions in the      
$T <\ T_c$ regime}

\author{Michele~Caselle$^a$, Luca~Castagnini$^b$, \speaker{Alessandra~Feo}$^{a}$,
Ferdinando~Gliozzi$^{a}$ and Marco~Panero$^{c}$\\
\llap{$^a$} Dipartimento di Fisica Teorica, Universit\`a di Torino and INFN Torino,\\
Via Giuria 1, 10125 Torino, Italy\\
\llap{$^b$} Institut f\"ur Theoretische Physik, Universit\"at Regensburg, \\
93040 Regensburg, Germany \\
\llap{$^c$}Department of Physics and Helsinki Institute of Physics, University of Helsinki, \\
FIN-00014 Helsinki, Finland \\
E-mail: \email{caselle@to.infn.it}, \email{luca.castagnini@physik.uni-regensburg.de},
\email{feo@to.infn.it}, \email{gliozzi@to.infn.it}, \email{marco.panero@helsinki.fi}
}

\abstract{We present Monte Carlo results for the thermodynamics of 
pure $SU(N)$ gauge theories with $N=2,...,6$ in $2+1$ dimensions.
We focus on the confined phase region $T<T_c$
and study thermodynamics variables such as the trace 
of the energy-momentum tensor, pressure, energy and entropy 
density using the integral method. We also
investigate scaling properties with $N$ of the different 
observables.
We compare our results with a gas of free glueballs 
and the bosonic string predictions for the Hagedorn spectrum. 
\vspace{4cm}
\begin{flushright}
DFTT 23/2010 \\
HIP-2010-28/TH
\end{flushright}   }

\FullConference{The XXVIII International Symposium on Lattice Field Theory, Lattice2010\\
		June 14-19, 2010\\
		Villasimius, Italy}

\begin{document}

\section{Introduction}
\label{intro}
The exploration of the phase diagram of QCD and its thermodynamics are 
challenging problems and central goals of lattice simulations at finite 
temperature and density. See \cite{kanaya} for a review.
In this work we present Monte Carlo results for the thermodynamics of 
$SU(N)$ gauge theories with number of colors, $N=2,...,6$, in $2+1$ dimensions.
These theories are closely related with those in $3+1$ dimensions and are 
more numerically feasible.
We focus on the confined phase 
$T<T_c$ and study thermodynamic
variables such as the trace of the energy-momentum tensor, pressure, 
energy and entropy density using the integral method. 
We also investigate scaling properties with $N$ of the different 
observables and compare our results with the predictions obtained assuming 
that the thermodynamics of the system could be described as 
a gas of free glueballs. We shall show that a relevant improvement 
in the comparison near the critical point is obtained including also higher 
orders in the glueball spectrum and assuming for these terms 
a bosonic string description.

\section{Thermodynamics on the lattice}
\label{method}
Before discussing the thermodynamics of $SU(N)$ lattice gauge theories 
in $2+1$ dimensions, we sketch some basic thermodynamics relations 
in the continuum. 
From the partition function $Z(T,V)$ we get the free energy density as,
\be
f = -\frac{T}{V} \log Z(T,V) \, ,
\ee
where $T$ is the temperature and $V$ is the spatial volume.
In the thermodynamic limit the pressure is related to the free energy
density as,
\be
p = - \lim_{V \to \infty} f \, .
\ee
In the following we will assume to have a large, homogeneous system, 
so that the pressure can be identified as minus the free energy.
Once the pressure is calculated as a function of the temperature $p(T)$,
the other thermodynamics variable are derived. For example,
the trace of the energy-momentum tensor $\varepsilon - 2 p$ is,
\be
\frac{\varepsilon - 2 p}{T^3} = T \frac{\partial}{\partial T} \left(\frac{p}{T^3}\right) \, .
\label{emt}
\ee
The energy density $\varepsilon =  T^2 \frac{\partial}{\partial T} (p/T) $ 
is then obtained by adding $2p/T^3$
to this result while the entropy is given by,
\be
s = \frac{\varepsilon + p}{T} = \frac{\partial p}{\partial T}  \, .
\ee

On the lattice the temperature and volume of the thermodynamic system are 
determined by the lattice size $N_\tau \times N_s^2$ and the lattice 
spacing $a$,
\be
V = (a N_s)^2 \, , \, \, \, \, \, \, \, \, \, \, \, \, \, \, \, 
T = \frac{1}{a N_\tau} \, . 
\label{nt}
\ee

In this work, we perform a non-perturbative study of $SU(N)$ Yang-Mills theories 
with $N = 2,3,4,5,6$ colors regularized on a finite lattice, with lattice spacing $a$, 
with $N_s$ points along the two space-like directions and $N_\tau$ points along 
the time-like direction. 
We use the Wilson action for a generic $SU(N)$ gauge group,
\be
S_W(U_\mu(x)) = \sum_P S(U_P) \, , 
\, \, \, \, \, \, \, \, \, \, \, \, \, \, \, 
S(U_P) = \beta \bigg(1 - \frac{1}{N} {\rm Re Tr} \, U_P \bigg) \, ,
\label{action}
\ee
where $P$ denotes one of the $3 N_\tau \times N_s^2$ plaquettes on the lattice 
and $U_P$ is the product of the $U_\mu$-matrices (with $\mu=0,1,2$) around  
each $1 \times 1 $ plaquette.
On the lattice the partition function is given by,
\be
Z = \int \prod_{x,\mu} d U_\mu(x) \exp(- S_W(U_\mu(x)) ) \, .
\ee
In the continuum limit eq. (\ref{action}) becomes the standard Yang-Mills 
action provided that,
\be
\beta = \frac{2 N}{a g^2} \, .
\label{beta}
\ee
In $2+1$ dimensions $g^2$ has dimensions of mass 
and sets the scale. 

Although in principle all thermodynamics variables can be calculated from the
free energy density, in practice, a direct computation of the partition function
on the lattice is not possible. Here we use the integral method of 
Refs. \cite{Boyd:1996bx,Engels:1990vr}, as in Ref. \cite{Panero:2009tv}. 
We first calculate the action, i.e.,
the derivative of the partition function with respect to the bare coupling 
$\beta$. 
Up to an integration 
constant, resulting from the lower integration limit 
$\beta_0$, the pressure is then obtained by integrating,
\be
\frac{p(\beta,N_\tau,N_s)}{T^3} = N_\tau^3 \int_{\beta_0}^\beta d \beta^\prime
\Delta S(\beta^\prime,N_\tau,N_s)
\label{energydensity}
\ee
where in 2+1 dimensions 
\be
\Delta S(\beta^\prime,N_\tau,N_s) =  3 \langle P_0 \rangle_\beta -
\langle P_s + 2 P_\tau \rangle_\beta \, .
\label{deltaS}
\ee
Here $P_{s,\tau}$ denote the expectation values of space-space,
space-time plaquettes, respectively and $P_0$ is the plaquette value on
symmetric lattices $N_s^3$.
Using eqs. (\ref{energydensity}) and (\ref{deltaS}) we can write the 
trace of the energy-momentum tensor (\ref{emt}) as,
\be
\frac{\varepsilon - 2 p}{T^3} = T \frac{\partial}{\partial T} \left(\frac{p}{T^3}\right) =
N_\tau^3 \Delta S\left(\beta\left(\frac{T}{T_c}\right),N_\tau,N_S \right) T \frac{d \beta}{d
 T} \, .
\label{EMT}
\ee
In order to obtain eq. (\ref{EMT}) as a function of $T/T_c$, where $T_c$ is 
the critical temperature of the continuum theory, we need to relate $T/T_c$ to 
$\beta$, for any value of $N$,
\be 
\beta = \beta(T/T_c) \, .
\label{betaT}
\ee
In practice,  
$ T \frac{d \beta}{d T}$ is determined through a parametric fit (similar to the 
one performed in \cite{Bialas:2008rk} for $SU(3)$) but for generic $N$. 
A good choice 
for $\beta_0$ \cite{Bialas:2008rk} is
$\beta_0 = \beta(T/T_c = 0.6) $,
after checking from the measurements that $N_\tau^3$ times the integrand in 
(\ref{EMT}) is negligible 
at this temperature. 

\section{$SU(N)$ gauge theories at large $N$: scaling properties}
Let us investigate the large-$N$ limit of eq. (\ref{betaT}) in $SU(N)$ 
gauge theories in $2+1$ dimensions. 
To do so we need to relate some dimensionless ratios that 
in this limit become constant \cite{Teper:2009uf,Lucini:2002wg,Liddle:2008kk}, say,
\be
\frac{m_0}{\sqrt{\sigma}} = 4.108(20) + \frac{c}{N^2} + ...\, , 
\, \, \, \, \, \, \, \, \, \, \, \, \, \, \,
\frac{T_c}{\sqrt{\sigma}} = 0.903(3) + \frac{0.88}{N^2} \,+ ... \, ,
\ee
where $c$ is a constant.
Here $\sqrt{\sigma}$ is the square root of the string tension at zero temperature 
in the continuum theory.

Considering also that, 
if we keep $g^2N$ fixed, 
$\beta$ scales as $N^2$ (\ref{beta}) and 
from \cite{Teper:2009uf,Lucini:2002wg} 
\be
\frac{\sqrt{\sigma}}{g^2 N} = 0.1975 - \frac{0.12}{N^2} + ...\, , 
\ee
we get 
\be
\sqrt{\sigma} = \frac{0.395 N^2}{a \beta} - \frac{0.24}{a \beta} \, + ... \, .
\ee
Combining these expressions we obtain the dependence of $\beta$ in 
terms of the temperature $T$.
To get $\beta(T)$, 
it is particularly convenient to set the temperature scale 
using the $\sqrt{\sigma}/T_c$ ratio. 
To the first order in $\beta$ we have
\be
\frac{T}{T_c} = \frac{T}{\sqrt{\sigma}} \frac{\sqrt{\sigma}}{T_c } = 
T \frac{a \beta}{(0.395 N^2 - 0.24)\left(0.903 + \frac{0.88}{N^2} \right)}
\ee
and using eq. (\ref{nt}) gives,
\be
\beta = N_\tau \frac{T}{T_c} \left(0.357 N^2 + 0.13 -0.211/N^2 \right) \, ,
\label{scaling0}
\ee
which for $N=3$ gives $\beta=0.34 $ (to be compared with the expression given in 
Bialas et al. \cite{Bialas:2008rk}, 
$\beta = 3.3 N_\tau \frac{T}{T_c} + 1.5 + O(1/N_\tau)$, 
which gives $\beta = 0.33$).
Combining Eq. (\ref{scaling0}) with the data from \cite{Liddle:2008kk}
to get the correction to the scaling in the large-$N$ limit we obtain,
\be
\beta = N_\tau \frac{T}{T_c} (0.357 N^2 + 0.13 -0.211/N^2) + (0.22 N^2 - 0.5) \, ,
\label{scaling}
\ee
which gives the dependence of $\beta$ on the temperature $T$ up to a first
order correction to be used in eq. (\ref{EMT}).

\section{Numerical results and discussions}

We are now ready to evaluate the trace energy-momentum tensor in eq. (\ref{EMT})
and check the validity of the scaling dependence in eq. (\ref{scaling})
by plotting the right hand side of eq. (\ref{EMT}) vs. $t \equiv \frac{T}{T_c}$.
This plot is expected not to be dependent on $N$ (and also on $N_\tau$).

The numerical simulations were performed using the Chroma library 
\cite{Edwards:2004sx}
plus our own programs (for $SU(2)$ and $SU(4)$).
We evaluated $\Delta S$ for $N_\tau=6$ (and for SU(2) and SU(3) also for $N_\tau=8$)
and spatial volumes $N_S^2$ such that the aspect ratio
was always $N_s/N_\tau \ge 8$. 
In agreement with Ref.~\cite{Bialas:2008rk}, 
we may safely assume that, 
in this temperature range, 
this condition is enough to eliminate finite size effects in the spatial directions.
A detailed description of our results and of the algorithms we used 
will be reported elsewhere~\cite{toappear}.

\begin{figure}
\begin{center}
\includegraphics[width=0.49\columnwidth]{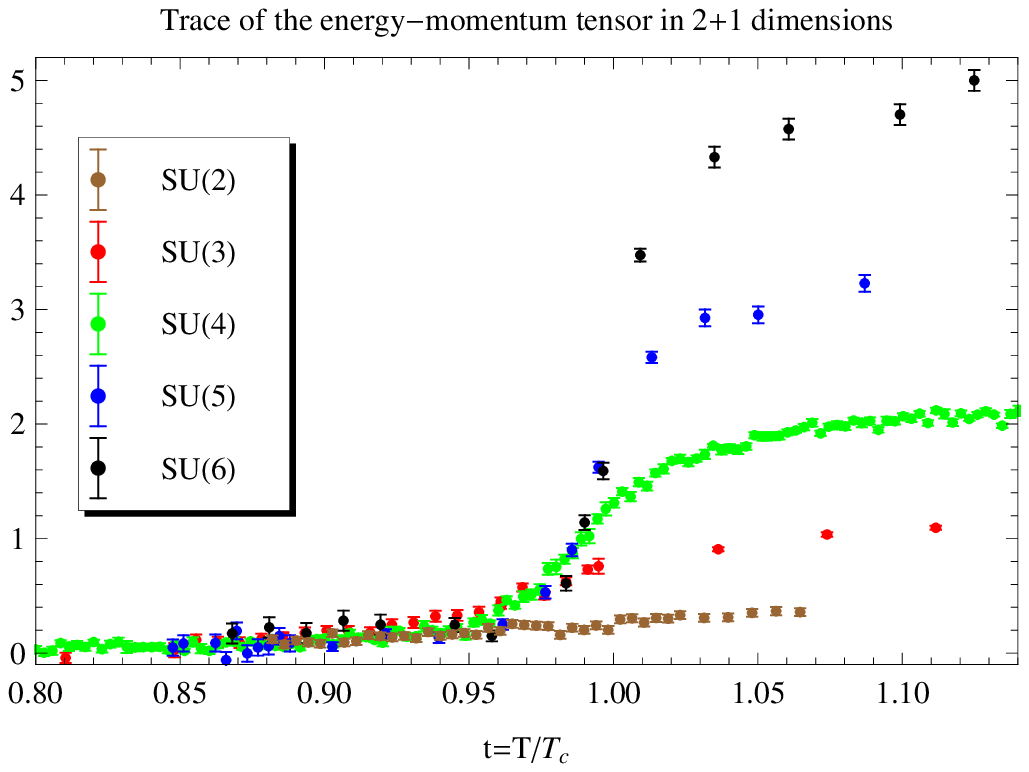}
\includegraphics[width=0.49\columnwidth]{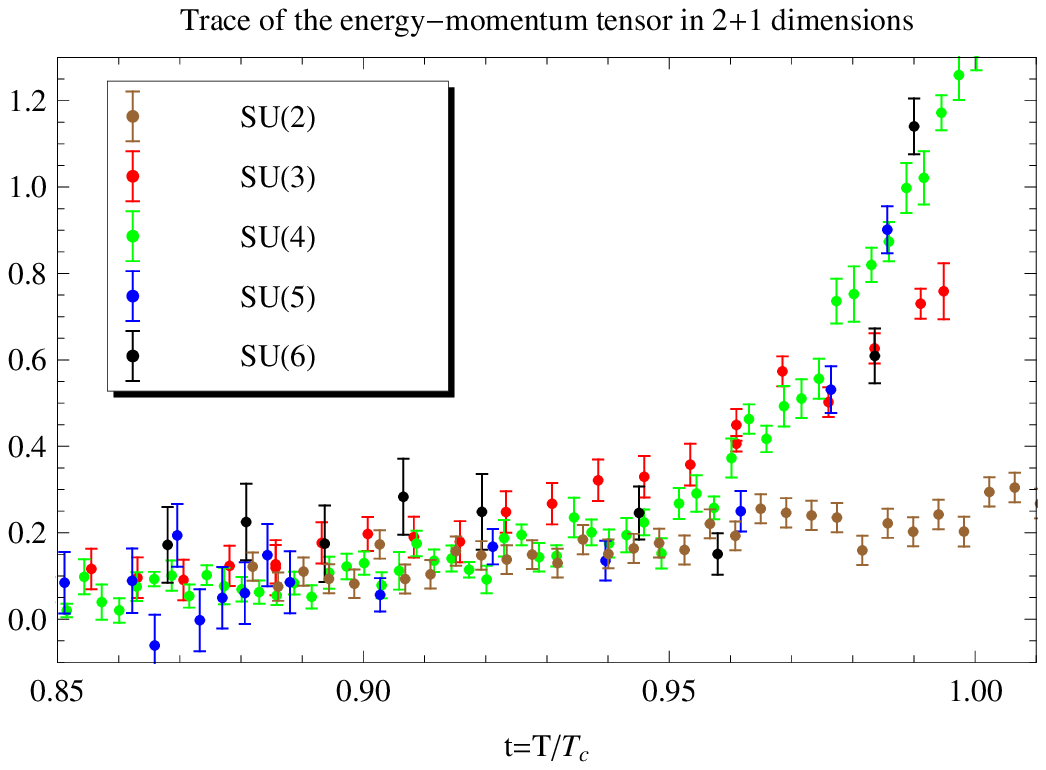}
\end{center}
\vspace{-0.5cm}
\caption{Left:The trace of the energy-momentum tensor vs. $t=T/T_c$ for 
$SU(N=2,3,4,5,6)$.
Right: Magnified view, of the same, in the low temperature region. }
\label{fig.trace}
\end{figure}

\begin{figure}
\begin{center}
\includegraphics[width=0.49\columnwidth]{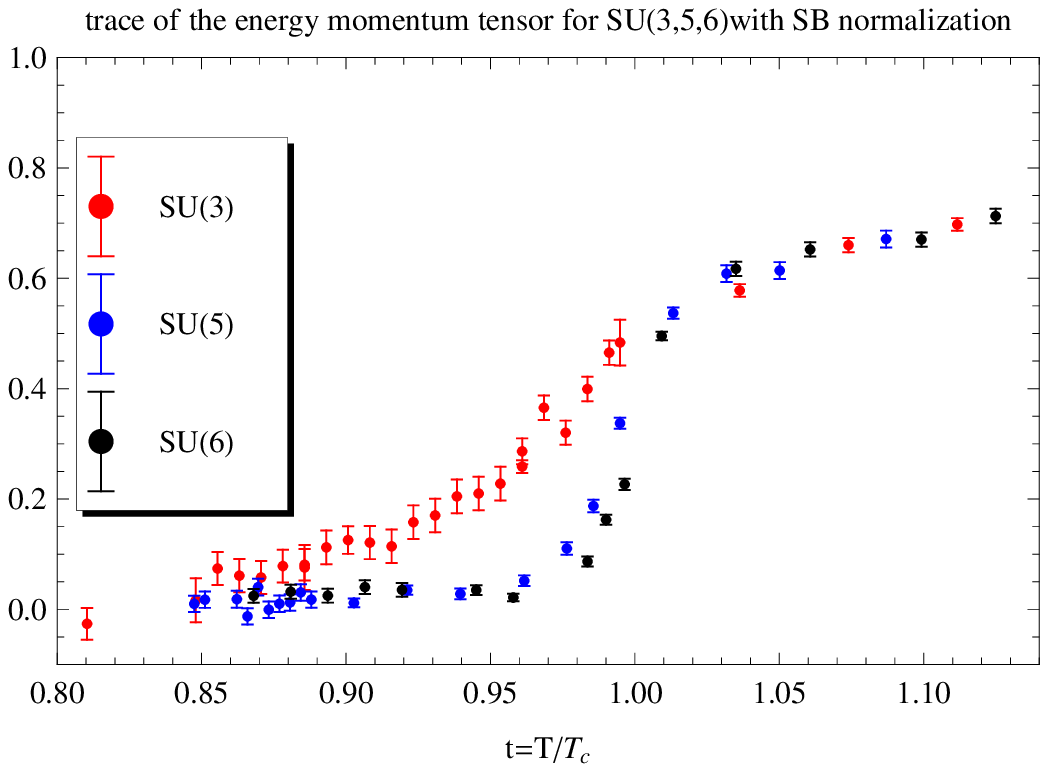}
\includegraphics[width=0.49\columnwidth]{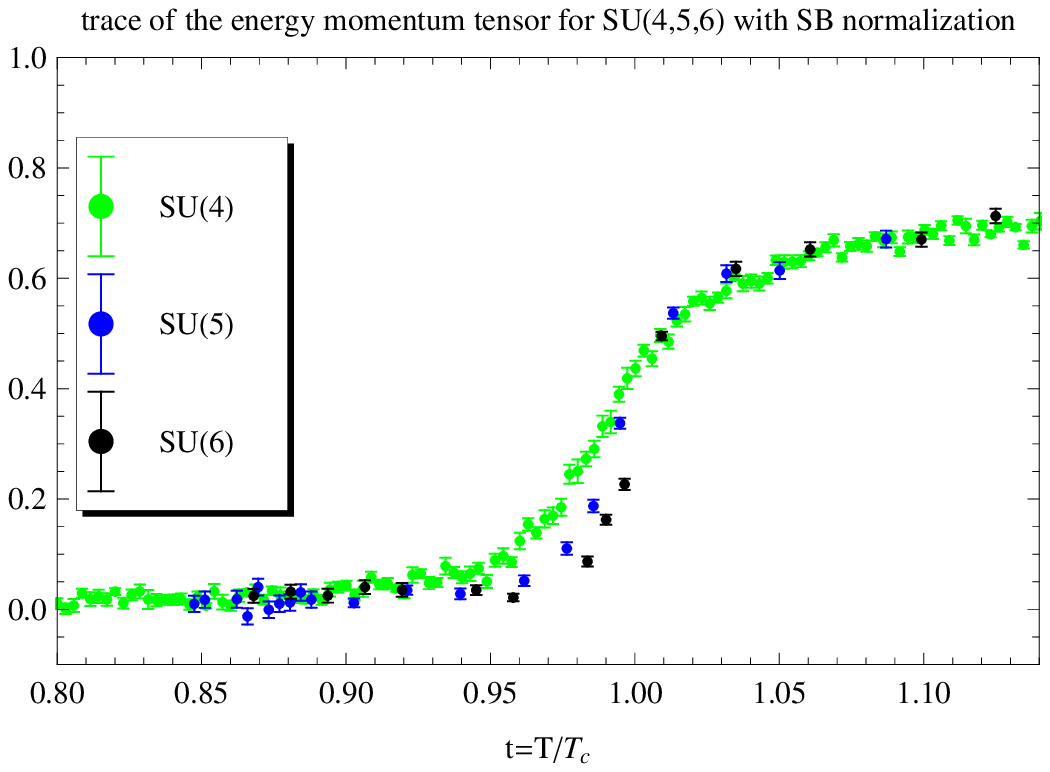}
\end{center}
\vspace{-0.5cm}
\caption{The trace of the energy-momentum tensor normalized to the lattice 
Stefan-Boltzmann (SB) limit vs. $t=T/T_c$.}
\label{fig.SB}
\end{figure}

We report in fig.~\ref{fig.trace} our estimates for the trace of the energy-momentum tensor.
The $SU(3)$ data are in perfect agreement with the one in 
Ref. \cite{Bialas:2008rk}. 
Below the critical temperature $T_c$ there is a good
scaling with $N$.
Above $T_c$, the different curves split up and they appear to be 
ordered according to the high temperature scaling law for 
the value of $N$. See fig. \ref{fig.SB} for the gauge groups 
$SU(3,4,5,6)$.

As mentioned in the introduction our main goal was to compare the $T<T_c$ data 
with a glueball gas model. We performed this comparison in three steps. 
First we assumed the gas to be dominated by the lowest glueball state, 
then we included all the glueballs below the two-particle threshold 
(for which very precise numerical estimates exist) and finally, following 
the suggestion of~\cite{Meyer:2009tq}, we try to
compare our data with the whole glueball spectrum, assuming for the glueballs 
an ansatz inspired by the effective bosonic string model. 
Details on the calculations can be
found in~\cite{toappear}.

The pressure associated with a single non-interacting, relativistic particle 
species of mass $m$ reads, 
\be
p = \frac{m T^2}{2 \pi} \sum \frac{1}{k^2} \exp\left(-k \frac{m}{T}\right) \left(1 + \frac{T}{k m}
\right)
\label{press}
\ee 
From Eq. (\ref{press}) we can reconstruct all thermodynamics quantities 
as explained earlier.
In particular, the trace of the energy-momentum tensor
(\ref{emt}) can be written as
\be
\frac{\varepsilon - 2 p}{T^3} = \frac{m^2}{2 \pi T^2} \sum 
\frac{1}{k} \exp\left(-k \frac{m}{T}\right) \, .
\label{new}
\ee

\begin{figure}
\begin{center}
\includegraphics[width=0.49\columnwidth]{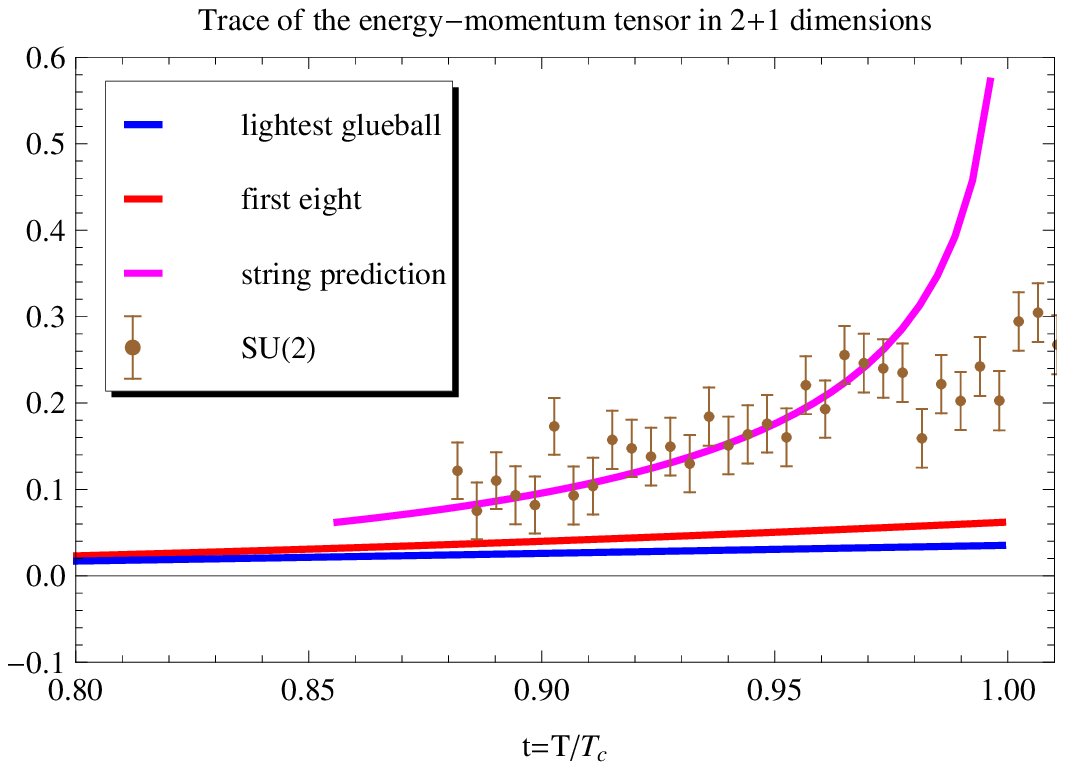}
\includegraphics[width=0.49\columnwidth]{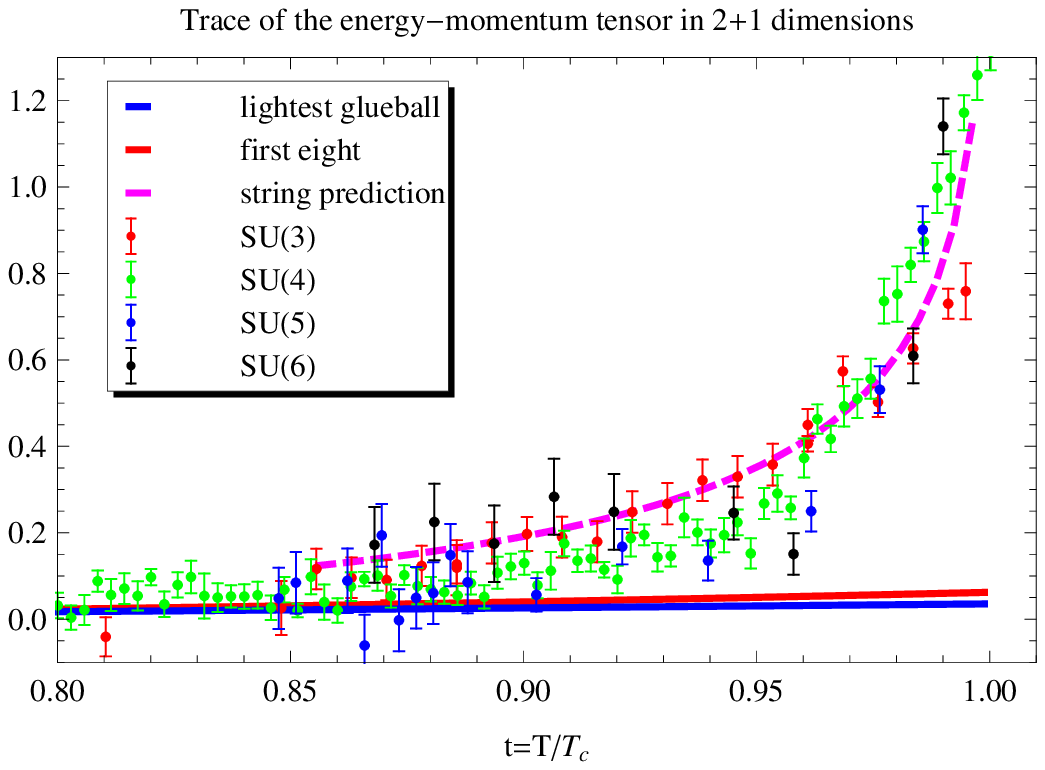}
\end{center}
\vspace{-0.5cm}
\caption{Comparison between contribution of the glueballs spectrum, the 
string predictions with respect to the trace energy-momentum tensor of $SU(2)$ (Left)
and $SU(N \ge 3)$ (Right).}
\label{fig.glue}
\end{figure}

This observable is particularly suited for this comparison since it can be calculated by
numerical simulations without being integrated over $\beta$ 
(see Eq. (\ref{EMT})). From the above equation we obtain:
\be
\Delta S = \frac{m^2 a^2}{2 \pi \beta N_\tau} \sum \frac{1}{k}  \exp\left(-k \frac{m}{T}\right)
 \, .
\label{deltaSfit}
\ee
For the first two stages of the comparison we used the numerical values of the glueball 
masses reported in \cite{Teper:1998te}. Given the precision of the data it is mandatory to keep into account scaling corrections in this comparison. 
The most effective way to do this is to rewrite $m/T$ as
\be
\frac{m}{T} = \frac{m}{\sqrt{\sigma}} \frac{\sqrt{\sigma}}{T} = 
\frac{m}{\sqrt{\sigma}} (aN_\tau \sqrt{\sigma}) \, , 
\ee
and then use the scaling functions reported in \cite{Teper:1998te}.
We write here these corrections explicitly in the $SU(3)$ case, for the lowest mass $m$ in the case of a lattice size $N_\tau=8$
(the generalization to any value of $N$ is straightforward). 
Using,
\be
a\sqrt{\sigma} = \frac{3.367(50)}{\beta} + \frac{4.1(1.7)}{\beta^2} 
+ \frac{46.5(11.0)}{\beta^3} \, ,
\label{msu3}
\ee
and  $\frac{m}{\sqrt{\sigma}} = 4.329(41) $ we obtain,
\be
 \frac{m}{T} = 8  \times 4.329 \times 
\left(\frac{3.367}{\beta} + \frac{4.1}{\beta^2} + \frac{46.5}{\beta^3}\right) \, .
\ee
Higher masses can be treated in the same way, using the data for the ratios 
$ m_i/\sqrt{\sigma}$ reported in
\cite{Teper:1998te}.
We compare the results of this analysis with the 
data for the trace of the energy-momentum tensor in fig. \ref{fig.glue} for $N=2$ and $N \ge 3$.
The blue and red lines correspond to the inclusion of the lightest 
mass and the first eight masses, respectively.
It is easy to see that these fits fail to reproduce the 
data and suggest the necessity of taking into account the 
full spectrum of glueballs
using for instance a string inspired ansatz.

To compare our results with the bosonic string predictions 
for the Hagedorn spectrum (in the same spirit as in 
Ref. \cite{Meyer:2009tq} in $4d$) we extended to arbitrary dimensions the computation of the density
of states of the closed bosonic string (following \cite{zwiebach}).

We found the following expression,
\be
\tilde \rho_{d-2}(M) = \left(\frac{\pi}{3}\right)^{d-1} \frac{1}{T_H} (d-2)^{\frac{d}{2} -1}
\left(\frac{T_H}{M}\right)^d e^{M/T_H} \, .
\label{string}
\ee
Inserting this expression in eq.~(\ref{new}) we found a remarkable agreement with the data, even 
in the region near the critical transition \cite{toappear}. 
This comparison is reported in fig.\ref{fig.glue}-left for $SU(2)$ and in 
fig. \ref{fig.glue}-right for $N \ge 3$.

We think that this type of analysis (which we plan to further improve in the future)
will give us the opportunity to test the string inspired 
glueball models (like for instance the Isgur-Paton one \cite{Isgur:1984bm}) and also to better understand  
the many non trivial features of effective string models which have been up to now addressed only looking at
observables related to the interquark potential or to the width of the flux tube.

\noindent{\large\bf Acknowledgements} \\
Numerical simulations were partially performed on the INFN Milano-Bicocca 
TURING cluster.
L.C. acknowledges partial support from Deutsche Forschungsgemeinschaft
(Sonderforschungsbereich/Transregio 55) and the European Union grant
238353 (ITN STRONGnet).
M.P. acknowledges financial support from the Academy of
Finland project 1134018.

\end{document}